# Differentiated roles of Lifshitz transition on thermodynamics and superconductivity in La$_{2-x}$Sr$_x$CuO$_4$

Yong Zhong[a,b,c†*], Zhuoyu Chen[a,c†§*], Su-Di Chen[a,c], Ke-Jun Xu[a,c], Makoto Hashimoto[d], Yu He[e], Shin-ichi Uchida[f], Donghui Lu[d], Sung-Kwan Mo[b], and Zhi-Xun Shen[a,c*]

[a]Stanford Institute for Materials and Energy Sciences, SLAC National Accelerator Laboratory, Menlo Park, CA 94025, USA

[b]Advanced Light Source, Lawrence Berkeley National Laboratory, Berkeley, CA 94720, USA

[c]Geballe Laboratory for Advanced Materials, Departments of Physics and Applied Physics, Stanford University, Stanford, CA 94305, USA

[d]Stanford Synchrotron Radiation Lightsource, SLAC National Accelerator Laboratory, Menlo Park, CA 94025, USA

[e]Department of Applied Physics, Yale University, New Haven, Connecticut 06511, USA

[f]Department of Physics, University of Tokyo, Tokyo 113-0033, Japan

[*]Correspondence to: ylzhong@stanford.edu, zychen@stanford.edu, zxshen@stanford.edu

[†]These authors contributed equally.

[§]Current address: Department of Physics, Southern University of Science and Technology, Shenzhen 518055, China (chenzhuoyu@sustech.edu.cn).

**Abstract**

**The effect of Lifshitz transition on thermodynamics and superconductivity in hole-doped cuprates has been heavily debated but remains an open question. In particular, an observed peak of electronic specific heat is proposed to originate from fluctuations of a putative quantum critical point $p^*$ (e.g. the termination of pseudogap at zero temperature), which is close to, but distinguishable from the Lifshitz transition in La-based cuprates. Here, we report an *in situ* angle-resolved photoemission spectroscopy study of three-dimensional Fermi surfaces in $La_{2-x}Sr_xCuO_4$ thin films ($x$ = 0.06 – 0.35). With accurate $k_z$ dispersion quantification, the Lifshitz transition is determined to happen within a finite range around $x$ = 0.21. Normal state electronic specific heat, calculated from spectroscopy-derived band parameters, agrees with previous thermodynamic microcalorimetry measurements. The account of the specific heat maximum by underlying band structures excludes the need for additionally dominant contribution from the quantum fluctuations at $p^*$. A $d$-wave superconducting gap smoothly across the Lifshitz transition demonstrates the insensitivity of superconductivity to the dramatic density of states enhancement.**

**Main Text**

**Introduction**

The underlying physics of the rich phase diagram in hole-doped cuprates remains an open question [1-4]. With the change of doping, a Lifshitz transition (LT) of the Fermi surface is expected when the chemical potential crosses the van Hove singularity (VHS), at which the Fermi surface transforms from a hole-like pocket around ($\pi$, $\pi$) to an electron-like pocket around (0, 0) [5]. Recent experimental progress, including the putative termination points of pseudogap [6], the presumed quantum critical points seen in electronic specific heat and Hall number [7], and the anomalously low superfluid density in electromagnetic response in over-doped regime [8,9], reveal unanticipated surprises. On the theoretical front, the presence of VHS was suggested to be important for the enhanced $T_C$ and phase fluctuations [10-12]. Two-dimensional Hubbard model simulations proposed that this transition has a strong influence on the phase diagram of hole-doped cuprates [13-15]. For certain band structure parameters, the VHS was associated with a critical point that changes the characters of superconductivity [14,15]. Given the richness of experimental phenomenology and theoretical postulations, a systematic study of the electronic structure across the LT is desirable.

Angle-resolved photoemission spectroscopy (ARPES) is a powerful tool to visualize Fermi surface in momentum space [16], yielding quantitative agreements with the results from quantum oscillation experiments [17-20]. LTs have been explored in different types of cuprates by ARPES measurements, such as $La_{2-x}Sr_xCuO_4$ (LSCO) [21], $YBa_2Cu_3O_{6+\delta}$ (YBCO) [22], $(Bi,Pb)_2Sr_2CuO_{6+\delta}$ (Bi2201) [23] and $Bi_2Sr_2CaCu_2O_{8+\delta}$ (Bi2212) [24]. Among them, LSCO is an ideal platform because the LT critical points (denoted as $p_{FS}$) in other cuprates are usually close to materials' doping solubility limit ($p_{FS}$ > 30% in YBCO, $p_{FS}$ > 35% in Bi2201, $p_{FS}$ > 29% in Bi2212, but $p_{FS}$ <

22% in LSCO). Nonetheless, the three dimensionality of LSCO electronic structure broadens the critical point $p_{FS}$ in two-dimensional case to a finite doping range [25], within which "electron-like" and "hole-like" features coexist and manifest at different $k_z$. Therefore, a detailed doping-dependent and $k_z$-dependent photoemission study is needed to dissect the LT in LSCO.

Here, we utilize a recently developed oxide molecular-beam epitaxy system *in situ* connected to a synchrotron ARPES endstation, for the synthesis of LSCO thin films with precise control of thickness and doping levels (See SI Appendix, fig. S1). Photon energy dependent ARPES measurements with improved spectral quality (See SI Appendix, fig. S2) reveal an accurate out-of-plane hopping parameter, determining that the LT occurs within a certain doping range. This is close to the doping of the normal-state electronic specific heat divergence and rapid Hall number change [26,27]. With the band parameters determined by fitting the electronic dispersion, we calculated the doping-dependent electronic specific heat and found a marked agreement with thermodynamic microcalorimetry experiment [26] in the vicinity of LT, providing a band theory explanation for the specific heat peak, without the need for the additional contribution from quantum fluctuations at a putative critical point $p^*$ (e.g. associated with the termination of pseudogap). Furthermore, we perform a systematic angle-dependent gap measurement which shows persistent *d*-wave symmetry across the LT, with no enhancement of the gap magnitude near the VHS. Our findings are incompatible with existing scenarios of high-$T_C$ superconductivity being associated with VHS or quantum critical point.

**Results**

Three-dimensional Fermi surface can be probed with in-plane momentum $\mathbf{k}_{//}$ (a combination of $k_x$ and $k_y$) and out-of-plane momentum $k_z$. In ARPES measurements, we access these momenta through the following relations

$$|\mathbf{k}_\parallel| = \sqrt{2m(h\nu - \phi - E_B)}\sin\theta/\hbar, \tag{1}$$

$$k_z = \sqrt{2m[(h\nu - \phi - E_B)\cos^2\theta + V_0]}/\hbar, \tag{2}$$

where *m* is the electron mass, $\phi$ is the work function, $E_B$ is the binding energy, $\theta$ is the polar angle, and $V_0$ is the inner potential. Firstly, as displayed in Fig. 1, we focus on the in-plane Fermi surface evolution at a selected photon energy $h\nu = 70\ eV$ (the selection of this photon energy will be discussed below). According to the Luttinger theorem, we calculate the doping levels via counting the filled states enclosed by the Fermi surface contours. In the following discussion, all the doping levels are determined by this method (SI Appendix, table S1 summarizes the information of samples). Secondly, we explore the out-of-plane Fermi surface by tuning the photon energies from 60 eV to 170 eV. A periodic $k_F$-$k_z$ dispersion on the $x = 0.22$ sample is shown in Fig. 2A (See SI Appendix, fig. S3 for more details). Inner potential $V_0 = 9$ eV is obtained by analyzing experimental data and considering the periodicity of Brillouin zones (BZ). To avoid the inconsistency from weaker spectral intensity around $E_F$ at some photon energies, we choose momentum $k_B$ (at binding energy $E_B = -20$ mV) as a fingerprint to quantify the $k_z$ dispersion. Sinusoidal-like $k_B$-$k_z$ relation is observed along (π, 0) to (-π, 0) direction on both $x = 0.22$ and 0.35 samples, in contrast to the negligible variation along (0, 0) to (π, π) direction (Fig. 2B). We use a three-dimensional tight-binding model to simulate the Fermi surface [28]

$$E_{3D}(k_x, k_y, k_z) = -\mu - 2t_1[\cos(k_x a) + \cos(k_y a)] - 4t_2 \cos(k_x a)\cos(k_y a)$$

$$-2t_3[\cos(2k_x a) + \cos(2k_y a)]$$

$$-2t_z \cos\left(\frac{k_x a}{2}\right)\cos\left(\frac{k_y a}{2}\right)\cos\left(\frac{k_z c}{2}\right)[\cos(k_x a) - \cos(k_y a)]^2 \quad (3)$$

Here $t_1$, $t_2$, and $t_3$ are the first, second, and third nearest neighbor hopping integrals between Cu sites. $t_z$ represents the interlayer hopping coefficient. $\mu$ is the chemical potential. $a$ is the in-plane lattice constant. $c/2$ is the distance between adjacent $CuO_2$ layers. In particular, the $\cos(k_x a/2)\cos(k_y a/2)[\cos(k_x a) - \cos(k_y a)]^2$ term accounts for the staggered stacking of neighboring $CuO_2$ planes in the body-centered tetragonal structure of LSCO. Considering the universality of constant nodal Fermi velocity in a wide doping range of LSCO [29], we choose $t_1 = 190$ (± 15) meV (determined by fitting the nodal dispersions from $E_F$ to – 30 meV on all measured samples) and $t_3/t_2 = -0.5$ for all samples, and tune $t_2$, $t_z$ and $\mu$ to depict the topography change of the Fermi surface (Table 1 and SI Appendix, fig. S4 summarize all the band parameters). The shift of $\mu$ is associated with the band filling effect from Sr dopants. The doping-dependent $t_2$ corresponds to a combined effect from chemical doping and electron correlation [30,31] and evolves smoothly across the LT. $t_z$ can be determined by fitting the experimental data via equation (3): $t_z = 0.030$ (± 0.004) $t_1$ on the $x = 0.22$ sample, and $t_z = 0.039$ (± 0.005) $t_1$ on the $x = 0.35$ sample. $t_z = 0.03 t_1$ is used to simulate the three-dimensional Fermi surfaces, as shown in the top right quadrant in each panel of Fig. 1. To quantify the LT regime accurately, we interpolate $t_2$ and $\mu$ linearly to obtain doping-dependent antinode's binding energy. This provides an estimated range of LT into $x = 0.206$ - 0.214 (the error bar for $x$ is 0.01), as shown in Fig. 2C. The three-dimensional Fermi surface for $x = 0.22$ is illustrated in Fig. 2D, where the red contour denotes the in-plane Fermi surface at $hv = 70\ eV$. We selected this photon energy in Fig. 1 because the corresponding $k_z$ ($\approx \pi / c$) locates at the midpoint between the largest and the smallest Fermi surfaces, making it the most representative for the study of LT and doping level characterization.

Having identified the LT in LSCO, we next examine its influence on the superconductivity. A typical Fermi surface map ($x = 0.22$) covering the first and second BZs is shown in Fig. 3A, in which we selected an antinodal cut in the first BZ and a nodal cut in the second BZ (for higher spectral intensity due to the matrix element effect in ARPES) as examples. The corresponding energy-momentum plots and energy distribution curves (EDC) are presented in Fig. 3B to E. While no gap opening is found at the node (Fig. 3C), there is a noticeable intensity depletion around $E_F$ for the antinode (Fig. 3B), implying an anisotropic gap structure. Quantitative analysis of gap symmetry in the superconducting state is shown in Fig. 4. As displayed in Fig. 4A to D, we obtained the gap size by fitting the symmetrized EDC at $k_F$ with respect to $E_F$, using formula $A(\omega) = -1/\pi$ Im$(1/(\omega-\Sigma(\omega)))$ with a constant background subtracted, where $\omega$ is the binding energy. The self-energy Σ(ω) term can be expressed as Σ(ω) = -iΓ+Δ²/ω, where Γ is the lifetime of quasiparticles, and Δ is the gap [32]. Gaussian convolution representing the instrumental resolution was applied to the spectral functions. For overdoped films ($x$ = 0.17, 0.22 and 0.24), the symmetrized EDC at antinode show two coherence peaks with a gap opening. Moving from the antinode towards the node, the gap size decreases, and finally a single peak forms at the node. Angle-dependent gap sizes are plotted in Fig. 4E to H. Canonical $d$-wave pairing symmetry is observed across the LT ($x$ = 0.17, 0.22 and 0.24). For $x$ = 0.26 and higher dopings, the gap is below the detectable limit. On the underdoped side ($x$ = 0.12), there is an obvious deviation of the antinodal gap from the $d$-wave extrapolation, indicating the dominant influence of pseudogap in the antinodal regime. This node-antinode dichotomy is a generic feature in underdoped cuprates [33,34].

**Discussion**

The established three dimensionality of Fermi surface is crucial for understanding the specific heat anomaly in LSCO. Thermodynamic measurements of Nd-LSCO display a divergence of the normal-state electronic specific heat in the vicinity of LT [26], which was seen as a key signature to the existence of quantum critical point $p^*$ inside the superconducting dome. Theoretically, electronic specific heat (usually referred to as Sommerfeld coefficient $\gamma$) is the temperature derivative of total quasiparticle entropy $S$ [35]

$$S = k_B \int [-f \ln f - (1-f) \ln (1-f)] \rho(E) dE. \tag{4}$$

Here $f$ is the Fermi-Dirac function, $k_B$ is the Boltzmann constant, and $\rho(E)$ represents the density of state (DOS) at binding energy $E$. Given that $\gamma$ is closely related to the DOS near $E_F$, we calculate the doping-dependent $\rho(E)$:

$$\rho(E) = (1/4\pi^3) \int d\mathbf{k}/\partial\varepsilon(\mathbf{k}), \tag{5}$$

where $d\mathbf{k}$ is the volume element in the momentum space. In the two-dimensional case, LT is associated with the divergence of DOS known as VHS, arising from the chemical potential crossing a saddle point at (π, 0), which also generates a significant anistropy of DOS along the Fermi surface (See SI Appendix, fig. S5 for details). In the LSCO system, three dimensionality reduces this divergence to a finite sharp peak. Below, we refer to this peak still as VHS for convenience.

In previous single crystal studies of LSCO, Nd-LSCO, and Eu-LSCO, VHS was considered insufficient to account for the electronic specific heat maximum [26,36]. In our analysis, we introduce doping-dependent $t_2$ and $t_3$ parameters extracted directly from ARPES spectra, incorporating electron-correlation effect caused by chemical doping. Considering the updated band parameters and the smaller $t_z$ we obtained, we revisit the relationship between VHS and the electronic specific heat by using equation (4). In order to make a direct comparison between thin film and single crystal, we benchmark our thin film data by performing ARPES measurement on a $x$ = 0.22 single crystal sample under the identical experimental setup. We find the same Luttinger volume (= 22%) and $t_z$ (= 0.032 ± 0.004 $t_1$) parameter as the results from $x$ = 0.22 thin film within experimental errors (See SI Appendix, fig. S6).

Figure 5A displays the calculated $\gamma$ based on the extracted band parameters from ARPES data. We compare the ARPES-derived $\gamma$ at 2 K with the data from thermodynamic measurements in LSCO and Zn-LSCO [36-38], revealing a marked consistency near the LT. Both the location (i.e. doping) and the magnitude of the specific heat peak agree between the ARPES derivation and thermodynamic measurements. The fact that such a simple approach captures the basic shape of the $\gamma$ curve is striking. This indicates that the topological transition of Fermi surface contributes dominantly to the specific heat peak in LSCO. Quantum critical fluctuations at $p^*$, if exist, possibly only play a minor role for the magnitude of specific heat near the LT in LSCO. Note that the ARPES measured band parameters (e.g. $t_2$) evolve smoothly across the LT (See SI Appendix, fig. S4), also indicating no sign of additional renormalization from quantum critical contributions. On the underdoped side (6% and 12%), doping-dependent Fermi arc length [40] is considered to calculate the $\gamma$ values (DOS is counted only within the Fermi arc regime), which also agrees with the thermodynamic data. Yet, the ARPES derived $\gamma$ in overdoped regime is roughly 30% smaller than the thermodynamic values. The temperature dependence of the $x = 0.22$ ARPES-derived $\gamma$ (See SI Appendix, fig. S7) shows a saturation below 2 K, which may explain the consistency between 2 K thermodynamic data [36] and extrapolated 0 K thermodynamic data [37,38] in LSCO. For Nd-LSCO and Eu-LSCO cases, additional $\gamma$ increase below 2K might signify additional effects from quantum criticality at $p^*$ [26], yet it is to be understood why the quantum critical point $p^*$ stem from other electronic orders (e.g. spin stripes peaking around 1/8 doping [39]) could coincide with the Fermi surface LT doping $p_{FS}$, unless such order makes the layers less coupled and thus more two dimensional. With more prominent $\gamma$ divergence, Nd-LSCO and Eu-LSCO have lower $T_C$ compared to LSCO, indicating a reversed correlation. We note that similar analysis between VHS and specific heat maximum has been reported in strontium ruthenates [41,42], and elastocaloric contribution to the entropy has been identified [43].

Previous specific heat analysis usually includes strong disorder scattering effect based on transport measurement [25,26,36]. Figure 5B summarizes the specific heat simulations under several $t_z$ parameters and scattering rate $\hbar/\tau$ [26]. However, the measured $\gamma$ of Zn-LSCO has no significant difference compared with LSCO [37,38] (also seen in Fig. 5), while disorder scattering of Zn-LSCO in transport measurements is apparently higher [44]. This casts questions on whether disorder effect in transport and thermodynamic measurements are identical. In our analysis of the ARPES-derived $\gamma$, we adopt the assumption that the effect of transport disorder is negligible in thermodynamic measurements, based on the observations of similar measured $\gamma$ values in LSCO with and without Zn-substitutions [37,38].

Except for the divergent behavior of $\gamma$, there are other experimental observations pointing to a putative quantum critical point $p^*$ in cuprates. One of them is the rapid Hall number change from $p$ to $1+p$ at the pseudogap endpoint in Nd-LSCO and YBCO [27,45], perhaps corresponding to a LT across the critical point. However, transport experiments display no sign of $p$ to $1+p$ transition over a wide range of doping in LSCO [46-49] and Bi2201 [50], which implies that the sharp change of Hall effect at $p^*$ is not a universal feature in cuprates. Furthermore, systematic doping-dependent ARPES antinodal data is not consistent with a second-order quantum phase transition [51].

Finally, we discuss the role of LT on superconductivity. Our results show that the VHS with significantly enhanced DOS has little effect on both the superconducting gap size and gap symmetry in the LSCO system. In Bi2201 and Bi2212, the LT is found experimentally coinciding with the overdoped endpoint of superconductivity [23,24]. This correlation has been considered as a possible interpretation for the disappearance of superconductivity in overdoped cuprates. Dynamical cluster approximation calculation [14] displayed a spin susceptibility transition from antiferromagnetic type peaked at $q = (\pi, \pi)$ to ferromagnetic type centered at $q = (0,0)$ across the critical point $p_{FS}$. In addition, renormalization group analysis [15] and quantum Monte Carlo simulation [52] both predicted the existence of ferromagnetic fluctuations at van Hove filling in two-dimensional Hubbard model, which has been observed recently in heavily overdoped $p$-type and $n$-type cuprates [53,54]. We have shown that LSCO system manifests robust $d$-wave superconductivity across the LT, and the transition point $p_{FS}$ is far away from the SC endpoint, incompatible to the theoretical prediction of triplet $p$-wave pairing beyond the LT. An alternative theoretical proposal to the reduced superconductivity is that the flat band dispersion near antinode makes the $d$-wave pairing sensitive to impurity concentrations that increases with $x$ [55-57].

In conclusion, by systematic ARPES measurements on LSCO thin films, we accurately determined the doping range where the LT occurs. Based on the similar $\gamma$ values between LSCO and more disordered Zn-LSCO that excludes disorder effect in specific heat measurements, we adopt a simple band model and found a marked consistency between $\gamma$ calculated from ARPES band parameters and that from thermodynamic measurements in the vicinity of LT. These results show that in the LSCO system, the proposed quantum criticality at $p^*$ is not required to account for the magnitude of the electronic specific heat peak. We further unambiguously show that the superconducting gap remains persistent $d$-wave symmetry with no obvious enhancement of magnitude on both sides of the LT. Our results reveal the differentiated roles of LT on thermodynamics and superconductivity in LSCO: the LT causes the thermodynamic specific heat peak but has almost no effect on superconducting gap size and symmetry.

**Materials and Methods**

La$_{2-x}$Sr$_x$CuO$_4$ thin films with $x$ = 0.06, 0.12, 0.17, 0.22, 0.26, 0.31, and 0.35 were synthesized on LaSrAlO$_4$ (001) substrates in a Veeco GEN930 MBE system with continuous supply of purified ozone. The growth process was monitored by in situ reflective high energy electron diffraction (RHEED). Metal sources are individually shuttered to realize atomic layer-by-layer control of the growth, ensuring an atomically flat surface without impurities for optimized ARPES measurements. All samples presented in this work were 13.3 nm thick (10 unit cells) with LaO termination. The growth was performed at 700 °C, under background pressure of $1 \times 10^{-5}$ Torr of purified ozone. After the growth, the sample was immediately transferred under ultra-high vacuum to the ARPES endstation at Stanford Synchrotron Radiation Lightsource beamline 5-2 equipped with a Scienta DA30 analyzer. The base pressure during the ARPES measurements was better than $3 \times 10^{-11}$ Torr. Linear-horizontally polarized light with $h\nu$ = 70 eV was used to map the Fermi surface and detect the gap structure. Measurement temperature was 9 K, and the corresponding energy resolution was 12 meV. The $k_z$ dispersion was studied within a photon energy range 60 – 150 eV and temperature at 150 K. Previous $Sr_3Ru_2O_7$ study [42] that found a quantitative agreement between the quasiparticle DOS and the electronic specific heat inspire us to use band parameters to derive and compare with specific heat data in LSCO. We use fitted band parameters from the entire Fermi surface to calculate the Fermi velocity and the associated DOS as described in the main text, since it is better constrained with lower error level than local extractions, especially for the antinode regime. More details about the growth and measurement are shown in Supplementary Material.

**Acknowledgments**

We thank S. A. Kivelson, D. H. Lee, T. P. Devereaux, D. J. Scalapino, P. J. Hirschfeld and E. Huang for stimulating discussions. ARPES experiments were performed at Beamline 5-2, Stanford Synchrotron Radiation Lightsource, SLAC National Accelerator Laboratory. This work was supported by the U.S. Department of Energy, Office of Science, Office of Basic Energy Sciences, Materials Sciences and Engineering Division, under Contract DE-AC02-76SF00515. The work at ALS, LBNL was supported by US DOE under contract No. DE-AC02-05CH11231.

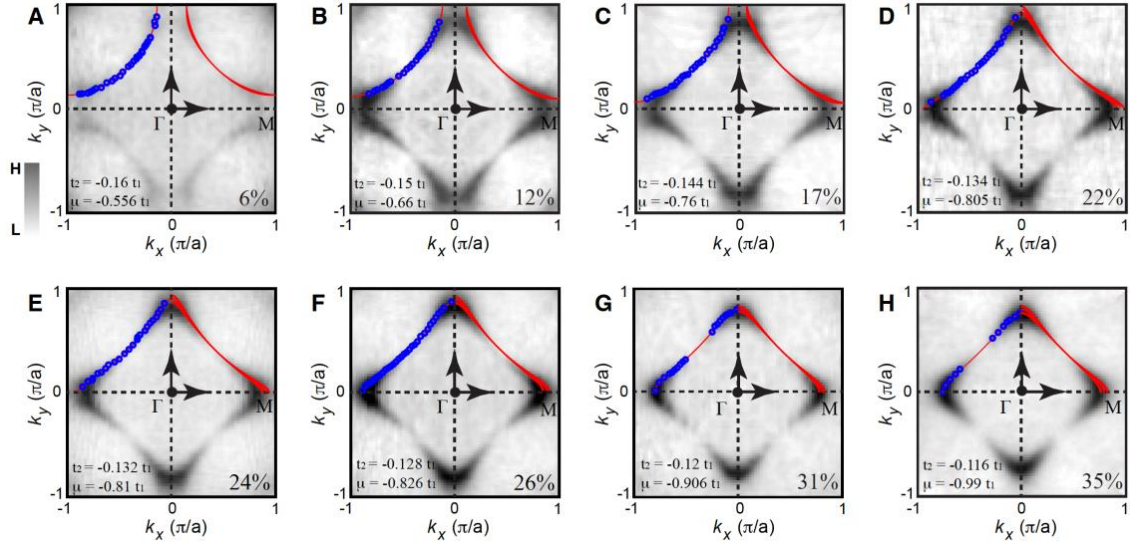

Fig. 1. **Three-dimensional Fermi surface projected onto $k_x$-$k_y$ plane.** (A-H) Doping-dependent in-plane Fermi surface maps. All the Fermi surfaces are integrated within the spectral weight window $E_F \pm 5$ meV ($x = 0.06$ sample with $E_F \pm 25$ meV due to the low intensity near $E_F$) and four-fold symmetrized in first Brillouin zone. Fermi momentum $k_F$ (blue circles) is determined by the peaks of the corresponding momentum distribution curves (MDC) at $E_F$. Red curves on top left quadrant of Brillouin zone are simulating in-plane Fermi surfaces at $k_z = \pi/c$ ($h\nu = 70eV$). Red curves on top right quadrant of Brillouin zone are simulating three-dimensional Fermi surfaces projected onto $k_x$-$k_y$ plane. We use $t_z = 0.03\ t_1$ for all simulations. Doping levels were determined by Luttinger theorem. Measurement temperature is 9 K. Details of band structure parameters $t_1$, $t_2$, $t_3$ and $\mu$ are listed in SI Appendix, table S1.

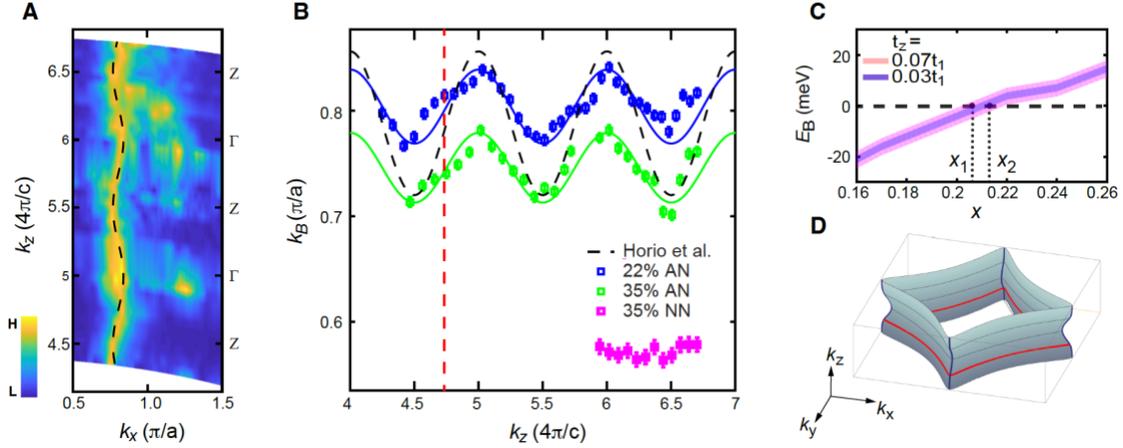

Fig. 2. **Three-dimensional Fermi surface projected onto $k_x$-$k_z$ plane.** (A) Out-of-plane Fermi surface maps along antinodal $(\pi, 0)$ direction for $x = 0.22$ sample. Photon energies are swept in the range 60 – 150 eV to cut different $k_z$ planes. (B) $k_B$ - $k_z$ dispersion along antinodal $(\pi, 0)$ direction for $x = 0.22$ sample (blue squares), and along (anti)nodal $(\pi, 0)/(\pi, \pi)$ directions for $x = 0.35$ sample (green and magenta squares). $k_B$ is the momentum located at the energy contour $E = -20$ mV. Error bars denote the uncertainty to obtain $k_B$. The blue and green lines represent the three-dimensional tight-binding simulations. The black dashed line comes from reference [26]. Red dashed line is the corresponding $k_z$ at photon energy $h\nu = 70 eV$. (C) Doping dependent binding energy $E_B$ at the antinodal momentum $k = (0.98\,\pi, 0.02\,\pi)$. Results from two interlayer hopping parameters ($t_z = 0.03\,t_1$, $0.07\,t_1$) are shown. Smaller $t_z$ narrows the Lifshitz transition regime into $x = 0.206 – 0.214$. Violet and pink areas represent the binding energy range for the two different $k_z$, respectively. $x_1$ and $x_2$ are the lower and upper bounds of the Lifshitz transition range. (D) Three-dimensional Fermi surface illustration of LSCO at $x = 0.22$. The red contour is the in-plane Fermi surface at $k_z = \pi/c$. Measurement temperature is 150 K.

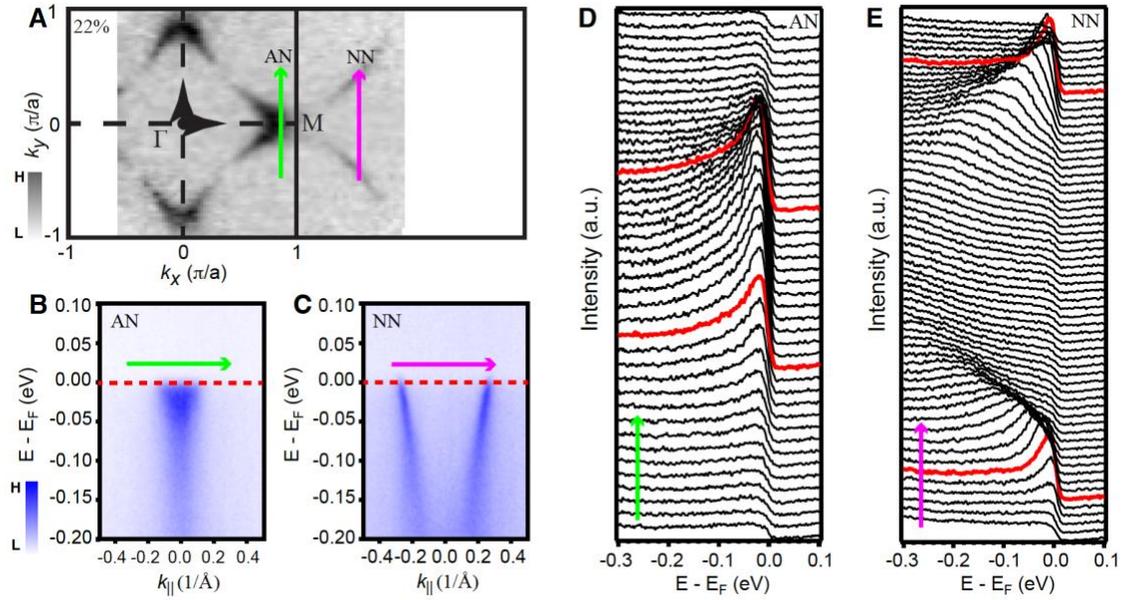

Fig. 3. **Low-energy electronic structure of LSCO at $x$ = 0.22.** (A) In-plane Fermi surface map covering first and second BZs. Due to the matrix element effect, we measure antinodal spectra in first Brillouin zone, and nodal spectra at second Brillouin zone. (B,C) Energy-momentum spectra along antinodal (green arrow) and nodal (magenta arrow) cuts. Red dashed line denotes the Fermi level. (D,E) The corresponding EDC curves along antinodal and nodal cuts labelled by green and magenta arrows respectively. Red curves are the EDCs at Fermi momenta $k_F$'s. $k_F$ in panel D is determined by the minimum gap criterion. Measurement temperature is 9 K.

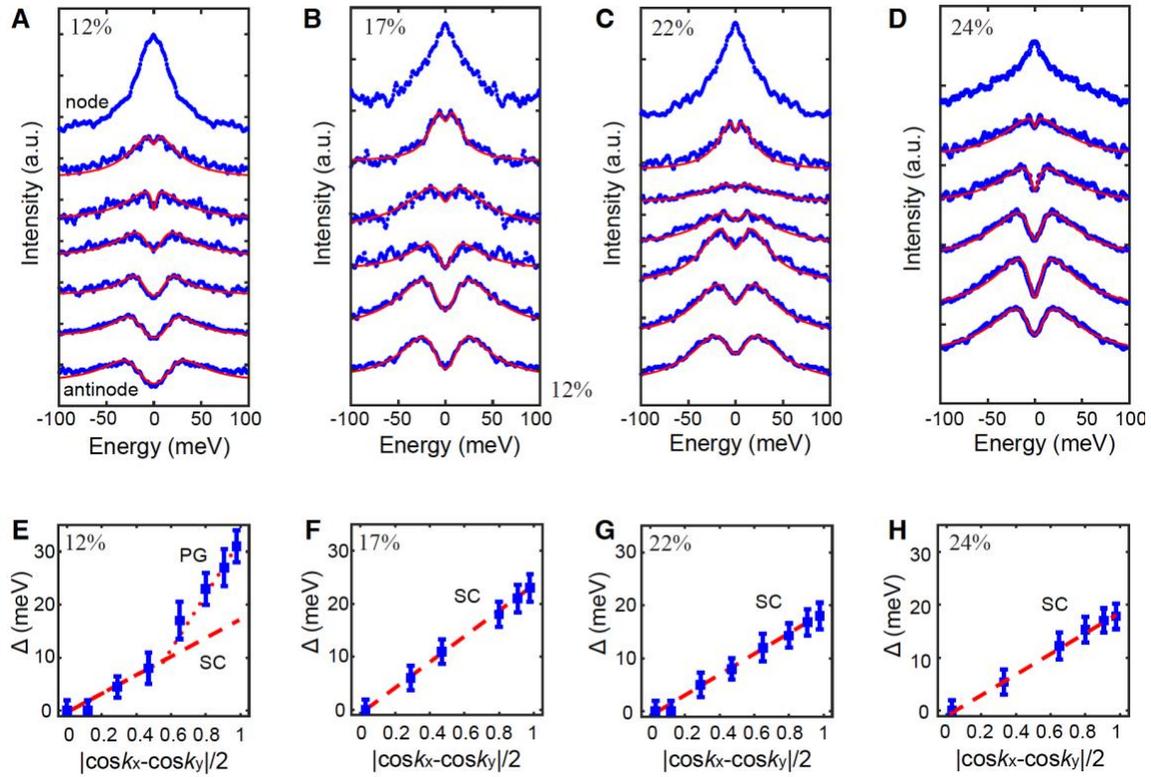

Fig. 4. **Superconducting gap symmetry.** (A-D) Angle-dependent symmetrized EDCs at $k_F$'s for x = 0.12, 0.17, 0.22 and 0.24 samples. The red lines are fittings to obtain the gap sizes. (E-H) Angle-dependent gap sizes for x = 0.12, 0.17, 0.22 and 0.24 samples. Dashed red lines are guides for the *d*-wave paring symmetry. Dotted red line denotes the dominant contribution of pseudogap around antinodal regime. Error bars reflect the fitting uncertainty in determining Δ. Measurement temperature is 9 K.

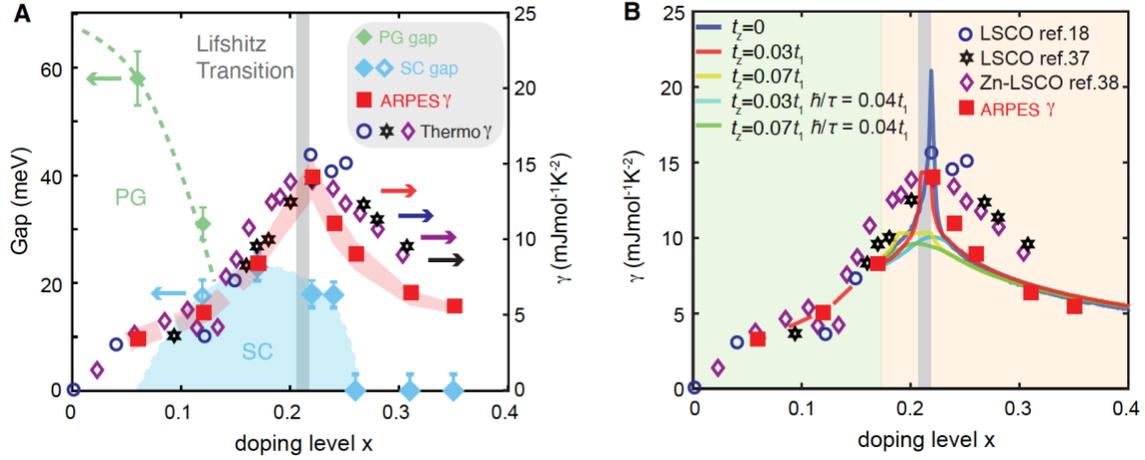

Fig. 5. **Doping-dependent electronic specific heat and phase diagram of LSCO.** (A) Phase diagram of LSCO. Blue solid diamonds are doping-dependent superconducting gaps. Blue open diamond corresponds to the extrapolated SC gap at 12% doping in Fig. 4E. Green diamonds correspond to the measured pseudogap in the underdoped regime. Error bars reflect the fitting uncertainty in determining the gaps. Grey vertical area represent the Lifshitz transition range same as the range between $x_1$ and $x_2$ in Fig. 2C. Red squares are electronic specific heat coefficients ($\gamma$) calculated at 2 K based on ARPES-derived band structure parameters (Fermi arc length is considered in the cases of x = 0.06 and 0.12). The shaded red curve represents the upper and lower bounds of the calculated $\gamma$ by considering the uncertainty of $t_1$ (175 – 205 meV) and the uncertainty of $t_z$. $t_z = 0.03t_1$ is used here in the calculation. Also plotted are electronic specific heat coefficients ($\gamma$) obtained by thermodynamic measurements: blue open circles from ref. 36 (measured at 2 K), black open stars from ref. 37 (extrapolated to 0 K), and purple open diamonds from ref. 38 (extrapolated to 0 K). (B) Calculated electronic specifc heat coefficients ($\gamma$) under different interlayer hopping parameters $t_z$ (0, 0.03 $t_1$ and 0.07$t_1$) and scattering rates $\hbar/\tau$ (0, 0.04$t_1$). $t_z = 0.07t_1$ is the hopping parameter used in ref. 25. $\hbar/\tau = 0.04t_1$ is the scattering rate used in ref. 26. The light yellow area indicates the regime in which we use underlying band structure to calculate specific heat coefficient $\gamma$. The light green area indicates the regime with the additional Fermi arc consideration.

**Table 1 Table of band structure parameters.**

| $x$ | $t_1$ (meV) | $t_2/t_1$ | $t_3/t_1$ | $t_z/t_1$ | $\mu/t_1$ |
|---|---|---|---|---|---|
| 6% | 190 | -0.16 | 0.08 | 0.03 | -0.556 |
| 12% | 190 | -0.15 | 0.075 | 0.03 | -0.66 |
| 17% | 190 | -0.144 | 0.072 | 0.03 | -0.76 |
| 22% | 190 | -0.134 | 0.067 | 0.03 | -0.805 |
| 24% | 190 | -0.132 | 0.066 | 0.03 | -0.81 |
| 26% | 190 | -0.128 | 0.064 | 0.03 | -0.826 |
| 31% | 190 | -0.12 | 0.06 | 0.03 | -0.906 |
| 35% | 190 | -0.116 | 0.058 | 0.03 | -0.99 |

Supplementary Information for

# Differentiated roles of Lifshitz transition on thermodynamics and superconductivity in La$_{2-x}$Sr$_x$CuO$_4$

Yong Zhong[1,2,3†*], Zhuoyu Chen[1,3†*], Su-Di Chen[1,3], Ke-Jun Xu[1,3], Makoto Hashimoto[4], Donghui Lu[4], Sung-Kwan Mo[2], S. Uchida[5] and Zhi-Xun Shen[1,3*]

*Correspondence to: ylzhong@stanford.edu, zychen@stanford.edu, zxshen@stanford.edu

†These authors contributed equally.

**This PDF file includes:**

Supplementary text

Figures S1 to S7

Tables S1

**Supplementary Text**

**Text S1. LSCO thin film preparation and ARPES measurements.**

The characterization of LSCO thin film is shown in Fig. S1. LaSrAlO$_4$ (001) substrate is chosen due to the relatively small lattice mismatch with LSCO. Before growth, the substrate is annealed in oxygen at 1000 °C with close contact to a LaAlO$_3$ substrate to ensure a AlO$_2$ termination. Real-time RHEED oscillations demonstrate an atomic layer-by-layer growth mode. The horizontal spacing between diffraction streaks of epitaxial LSCO film is used to calibrate the size of Brillouin zone in ARPES measurements. Wide-angle X-ray diffraction measurements confirm single phase in our films. Clear Kiessig fringers around main peaks reflect the high quality of the sample. AFM image shows atomically smooth surface.

In ARPES measurements, the chemical potential was calibrated by a polycrystalline gold film connected to the sample stage. Linearly polarized light with $h\nu$ = 70 eV was used to map the evolution of in-plane Fermi surfaces and explore the superconducting gap symmetry. Doping-dependent EDC curves at antinode/node, and MDC curves at node are displayed in Fig. S2. Ageing effect is also explored by warming up the sample ($x$ = 0.26) to 50 K and then cooled down to 9 K. EDC curves are recovered after this cycle, indicating negligible ageing effect in our films. Photo-energy dependent ARPES measurements were carried out to access the information of out-of-plane Fermi surface. Antinodal MDCs are used to extract the momenta $k_B$'s, as shown in Fig. S3.

**Text S2. Doping level determination.**

We use Luttinger theorem to get the doping levels in LSCO thin films. The error bars are ± 1.5% estimated by the uncertainty of $k_F$'s determination. Meanwhile, we also measure the elemental fluxes by quartz crystal microbalance (QCM) during growth. Both methods show a similar trend of $x$, as shown in Table S1. QCM displays a large deviation in heavily overdoped regime, indicating the Sr solution limit in LSCO thin film. In addition, we benchmark the doping level in $x$ = 0.22 thin film by performing ARPES measurement with the identical experimental setup on a $x$ = 0.22 single-crystal sample. Similar Luttinger volume, EDC curves, superconducting gap and interlayer hopping parameter $k_z$ are found (Fig. S6).

**Text S3. Tight-binding model calculations and simulations.**

We use three-dimensional tight binding model to calculate the doping-dependent DOS and compare it with thermodynamic properties. Doping-dependent band structure parameters determined by ARPES measurements are plotted in Fig. S4. Based on the parameters, angle-dependent DOS can be obtained in Fig. S5. Comparing with the underdoped ($x$ = 0.12) and heavily overdoped ($x$ = 0.35) samples, the Fermi velocity (then the DOS) near antinode in the Lifshitz transition range ($x$ = 0.22) changes dramatically. In contrast, the DOS near node seems no substantial variation. This means that the antinodal regime contributes mostly to the thermodynamic properties. Temperature-dependent electronic specific heat is shown in Fig. S7.

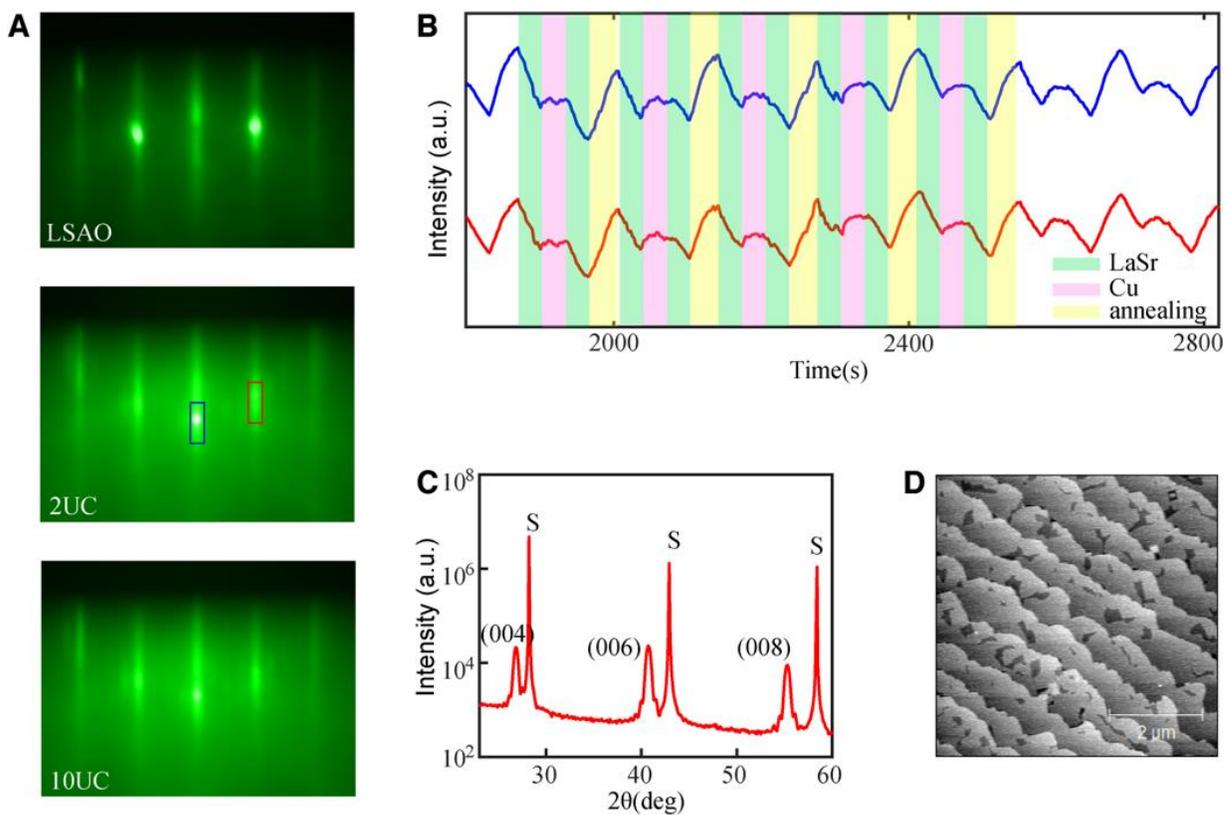

Fig. S1 **Characterization of LSCO thin films.** (A) RHEED patterns of LaSrAlO$_4$ (001) substrate (top), 2UC LSCO thin film (middle), and 10UC LSCO thin film (bottom). (B) Real-time RHEED intensity oscillations recorded during growth process. Blue and red curves are intensity of (00) and (01) streaks labelled in the middle panel of (A). (C) Wide-angle X-ray diffraction scan of a 10UC LSCO thin film. (D) Atomic force microscopy topography of thin film.

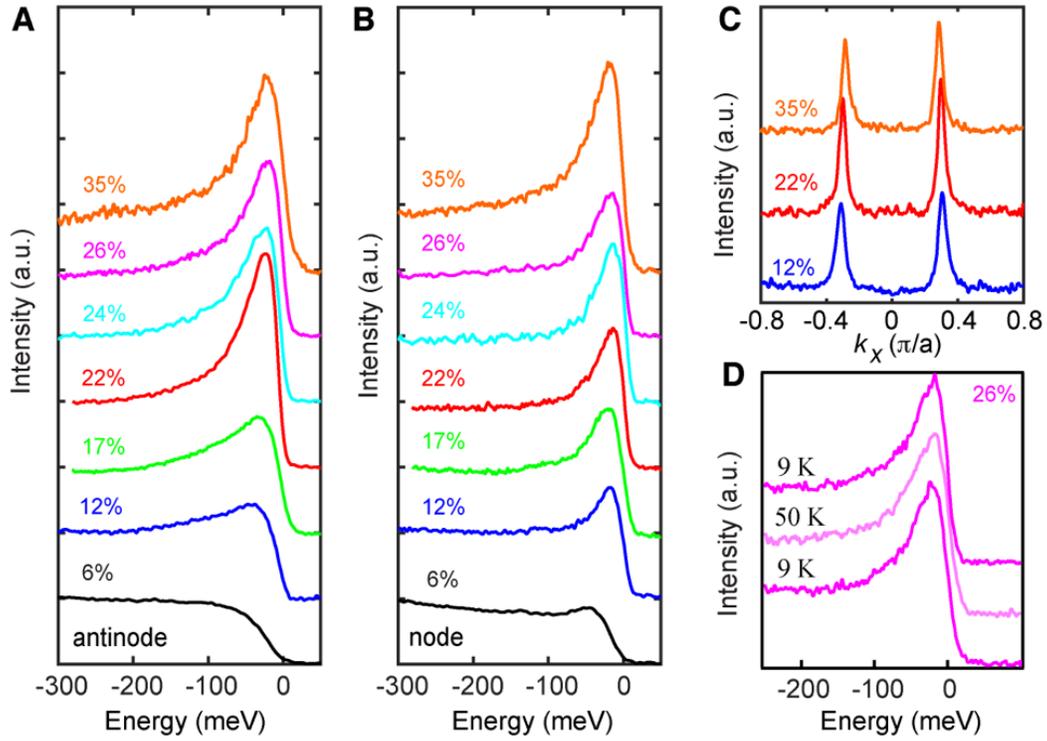

Fig. S2. **Doping-dependent EDC and MDC curves.** (A-B) Doping-dependent EDC curves at antinode/node. (C) Doping-dependent MDC curves along nodal direction. (D) Negligible ageing effect in LSCO thin films.

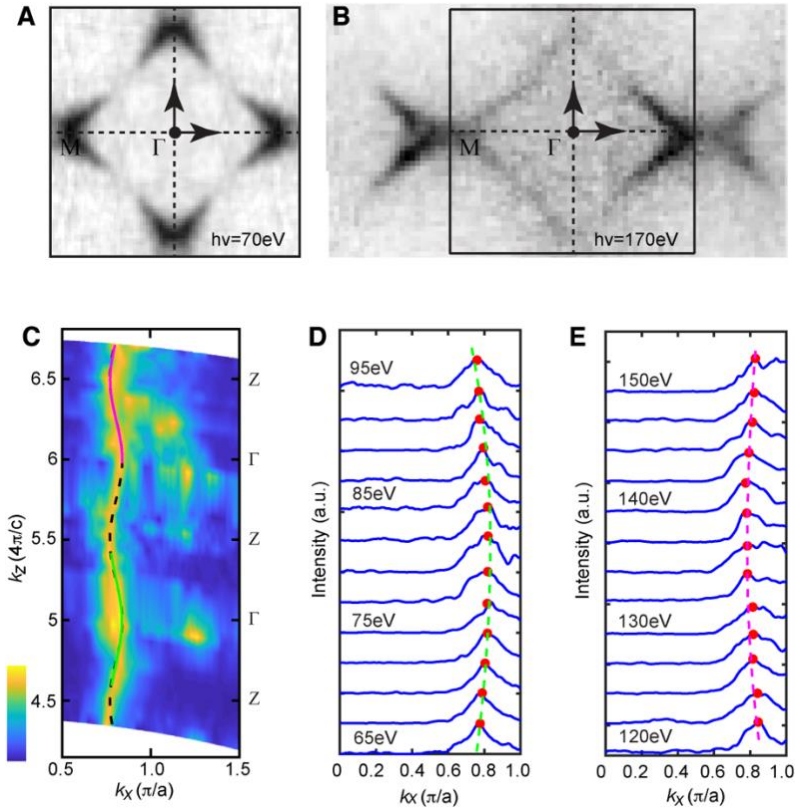

Fig. S3. **Photon energy dependent ARPES measurements.** (A-B) In-plane Fermi surfaces at photon energies $h\nu = 70eV$, 170eV. (C) Out-of-plane Fermi surface maps along antinodal $(\pi, 0)$ direction for *x* = 0.22 sample. (D-E) MDC curves at momenta $k_B$'s in two energy ranges, 65 – 95 eV and 120 – 150 eV.

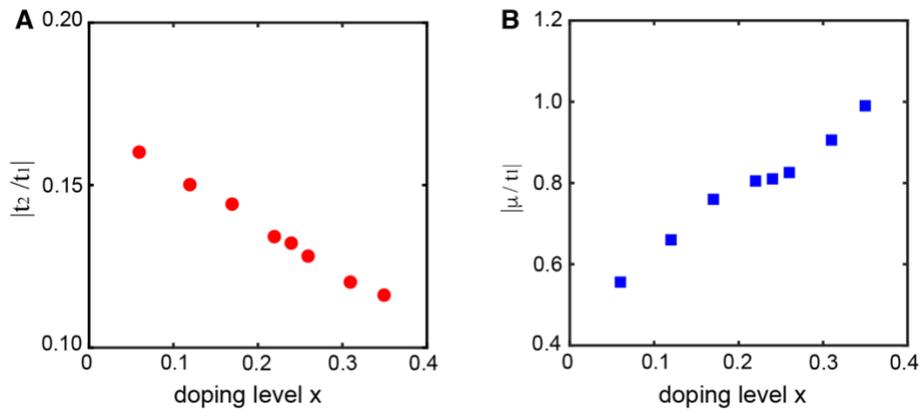

Fig. S4. **Doping-dependent band parameters.** (A-B) Doping-dependent $t_2$ and $\mu$.

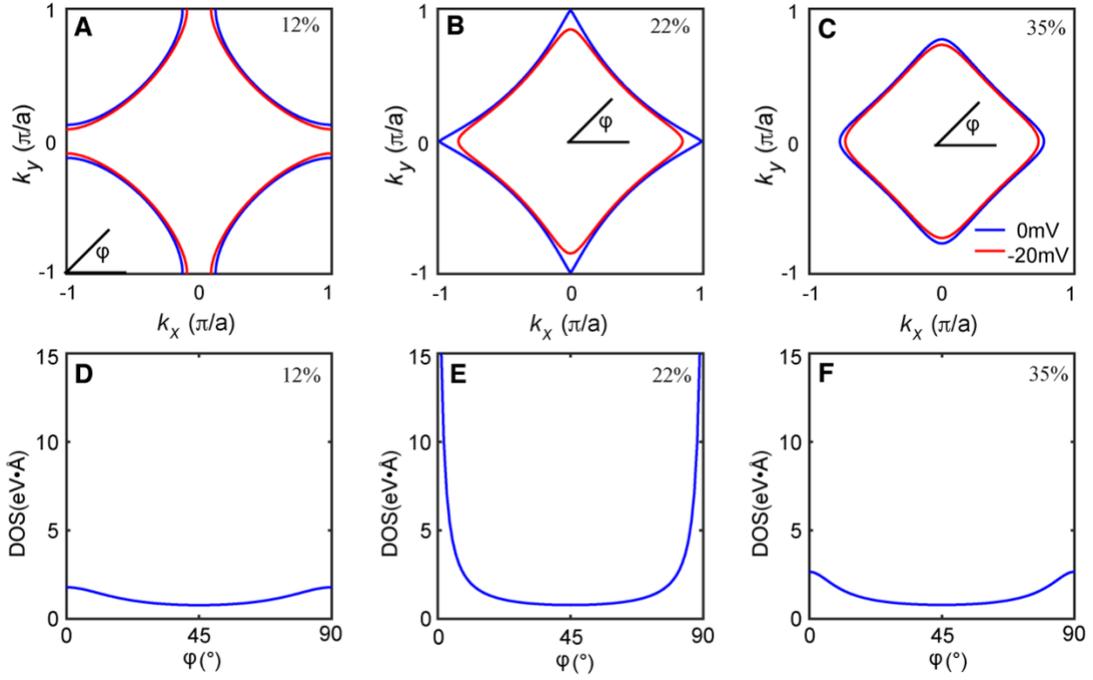

Fig. S5. **Angle-dependent density of states (DOS) in k-space.** (A-C) Energy contours (E = 0 meV, 20 meV) on *x* = 0.12, 0.22 and 0.35 samples. Azimuth angle $\varphi$ is defined in the figures. (D-F) The corresponding angle-dependent DOS in momentum space.

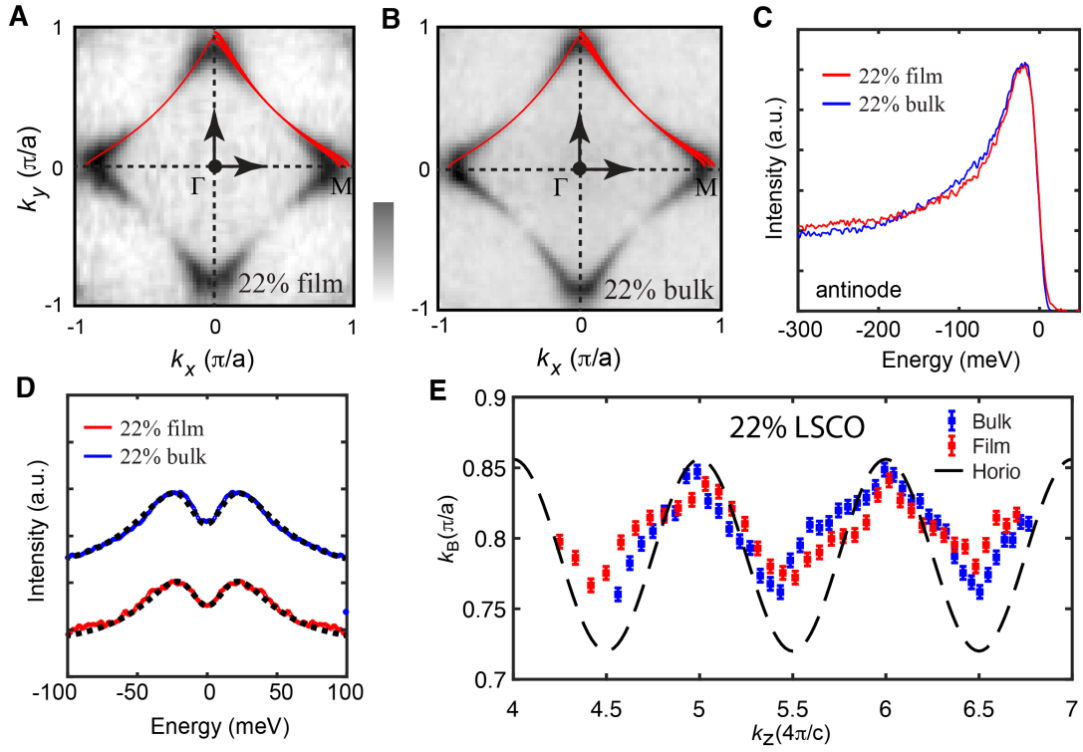

Fig. S6. **Comparison of *x* = 0.22 thin film and *x* = 0.22 single crystal.** (A-B) Fermi surface maps on *x* = 0.22 thin film and *x* = 0.22 single crystal. (C) Comparison of EDCs at antinode. (D) Comparison of symmetrized EDCs at antinode. (E) Similar interlayer hopping $k_z$.

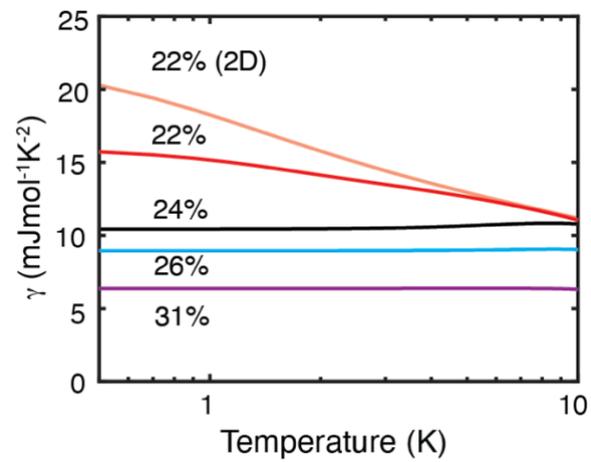

Fig. S7. **Temperature-dependent $\gamma$**. Temperature-dependent $\gamma$ is calculated in $x$ = 0.22, 0.24, 0.26 and 0.31 samples under $t_z$ = 0.03 $t_1$. In addition, we also show the two-dimensional case ($t_z$ = 0) in $x$ = 0.22.

**Table S1 Table of elemental fluxes for LSCO samples.**

| La ($2 \times 10^{13} atoms/cm^2*s$) | Sr ($2 \times 10^{13} atoms/cm^2*s$) | QCM $x$ | Luttinger theorem $x$ |
|---|---|---|---|
| 1.01 | 0.02 | 3.9% | 6% |
| 1.01 | 0.05 | 9.4% | 12% |
| 1.01 | 0.08 | 14.6% | 17% |
| 1.01 | 0.1 | 18.0% | 22% |
| 1.01 | 0.12 | 21.2% | 26% |
| 1.01 | 0.22 | 39.2% | 31% |
| 1.01 | 0.32 | 48.1% | 35% |